\documentclass[prl,english,twocolumn,amsfonts,amssymb,floats,showpacs,aps]{revtex4-1}

\usepackage[final,dvips]{epsfig}
\usepackage{amsmath,amsfonts}
\usepackage{bbm}
\usepackage{float}
\usepackage[caption = false]{subfig}
\usepackage{graphicx}
\usepackage{color}
\usepackage{hyperref}
\usepackage{multibib}

\newcommand{\jp}{J} % Inter-chain coupling Jp
\newcommand{\jn}{K} % Intra-chain NN Heisenberg coupling 
\newcommand{\dm}{D} % Intra-chain DM interaction
\newcommand{\gm}{\Gamma} % Bond anisotropy
\newcommand{\hy}{h_y} % External field

\newcommand{\ii}{\iota} % Math symbol for \sqrt(-1)

\begin{document}

\title[]
{
Topological quantum paramagnet in a quantum spin ladder
}
\author{Darshan G. Joshi}
\email[]{d.joshi@fkf.mpg.de}
\affiliation{Max-Planck-Institute for Solid State Research, D-70569 Stuttgart, Germany}

\author{Andreas P. Schnyder}
\email[]{a.schnyder@fkf.mpg.de}
\affiliation{Max-Planck-Institute for Solid State Research, D-70569 Stuttgart, Germany}

\date{\today}

\begin{abstract}
It has recently been found that bosonic excitations of ordered media, such as phonons or spinons, can exhibit topologically nontrivial band structures.
Of particular interest are magnon and triplon excitations in quantum magnets, as they can easily be manipulated by an applied field.
Here we study triplon excitations in an S=1/2 quantum spin ladder and show that they exhibit nontrivial topology,
even in the quantum-disordered paramagnetic phase.
Our analysis reveals that the paramagnetic phase actually consists of two separate regions with
topologically distinct triplon excitations.   
We demonstrate that the topological transition between these two  regions can be tuned by an external magnetic field.
The winding number that characterizes the topology of the triplons   is derived and evaluated. %{\clr 
By the bulk-boundary correspondence,
we find that the non-zero winding number implies the presence of localized triplon end states. %}
Experimental signatures and possible physical realizations of the topological paramagnetic phase are discussed.
\end{abstract}

%\pacs{ }

\maketitle

%%%%%%%%%%%%%%%%%%%%%%%%%%%%%%%%%%%%%%%%%%%%%%%%%%%%%%%%%%%%%%%%%%%%%%%%%
%%%%%%%%%%%%%%%%%%%%%%%%%%%%%%%%%%%%%%%%%%%%%%%%%%%%%%%%%%%%%%%%%%%%%%%%%

The last decade has witnessed tremendous progress 
in understanding and classifying topological band structures of fermions
\cite{hasan_kane,qi_zhang,classif1,classif2}. 
Soon after the discovery of fermionic topological insulators~\cite{koenig_qshe_science_07,hsieh_hasan_nature_09}, it was recognized that bosonic excitations of ordered media can as well exhibit
topologically nontrivial bands~\cite{shindou_murakami_PRB_13,zhang_li_magnonics_PRB_13,ss_hall,peano_marquardt_PRX_15,joannopoulos_review_photonics}.
Such bosonic topological bands have been observed not long ago for photons in dielectric superlattices~\cite{wang_joannopoulos_photonics}. %{\clr 
Theoretical proposals of topological states in 
polaritonic systems have been made \cite{polr_1d_th,jacqmin_bloch_polaritons_14,polr_bec}, some of which have been observed experimentally \cite{polr_1d_exp}. %}
Besides these examples, bosonic band structures are also realized by elementary excitations of quantum spin systems, e.g., by magnons 
in (anti)ferromagnets or by triplons in dimerized quantum magnets.

The study of these collective spin excitations 
is enjoying growing interest, due to potential applications for magnonic devices and spintronics~\cite{magnonics}.
Because magnetic excitations 
are charge neutral, they are weakly interacting, and therefore exhibit good coherence and
support nearly dissipationless spin transport. Moreover, the properties of spin excitations are easily tunable 
by magnetic fields of moderate strength, as the magnetic interaction scale is in most cases relatively 
small.  Of particular interest are magnetic excitations with   nontrivial band structure topology, since they exhibit protected magnon or triplon edges states. 
This was recently studied for triplons  in the ordered phase of the Shastry-Sutherland model~\cite{ss_hall,ss_hall_exp,kps_ss} and
for magnons in an ordered pyrochlore antiferromagnet~\cite{weyl_magnon} 
 as well as in an ordered honeycomb ferromagnet~\cite{sk_kim}.
However, the development of a comprehensive  
topological band theory for magnetic excitations
is still in its infancy. Specifically, it has remained unclear whether topological spin excitations can
 exist also in quantum disordered paramagnets. 
%which does not break any symmetry of the parent  spin Hamiltonian. 

\begin{figure}[b]
\centering
\includegraphics[width=0.4\textwidth]{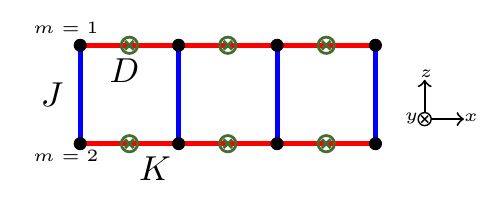}
\caption{
%{\clr [*change green labels and $y$ coordinate axis*]}
Schematic representation of the exchange interactions in the quantum spin ladder
described by Eq.~\eqref{eq:model}. The spins are shown as black circles, 
blue lines represent intra-dimer exchange ($\jp$),
and red lines inter-dimer interactions ($\jn$).
The DM interaction ($\dm$), indicated in green, points in the $y$-direction into the plane of the ladder. In addition, the model exhibits an even-parity spin-anisotropic  interaction ($\gm$), which arises along with the odd-parity DM interaction due to spin-orbit coupling. 
}
\label{fig:model}
\end{figure}

In this paper, we address this question by considering, as a prototypical example, the paramagnetic phase of an S=1/2 quantum spin
ladder with strong spin-orbit coupling. The considered spin ladder model describes a large class of 
well studied compounds, called coupled-dimer magnets~\cite{spin_ladder_review_Dagotto618}, 
which have two antiferromagnetically coupled spins per crystallographic unit cell (see Fig.~\ref{fig:model}).
Due to the strong antiferromagnetic exchange coupling within each unit cell, the magnetic ground state of these compounds is a dimer quantum paramagnet, where the two spins in each unit cell form a spin singlet. 
Examples of S=1/2 spin ladder materials include 
NaV$_2$O$_5$~\cite{nvo5}, Bi(Cu$_{1-x}$Zn$_x$)$_2$PO$_6$~\cite{bpo6}, and the cuprates 
SrCu$_2$O$_3$ \cite{sco3}, CaCu$_2$O$_3$ \cite{cco3},  BiCu$_2$PO$_6$ \cite{bcp_exp}, and LaCuO$_{2.5}$ \cite{lco2}.
%SrCu$_2$(BO$_3$)$_2$~\cite{kageyama_PRL_99}
Particularly interesting among these is BiCu$_2$PO$_6$, since it exhibits strong spin-orbit couplings,
which lead to  spin-anisotropic even-parity exchange couplings as well as
 odd-parity Dzyaloshinskii-Moriya (DM) interactions. 
As we will show, the latter gives rise to topologically nontrivial triplon exctiations.

 %%%%%%%%%%%%%%%%%%%%%%%%%%%%%
%%%%%%%%%%%%%%%%%%%%%%%%%%%%%%

%%%%%
\begin{figure*}[t]
\centering
\includegraphics[width=1.0\textwidth]{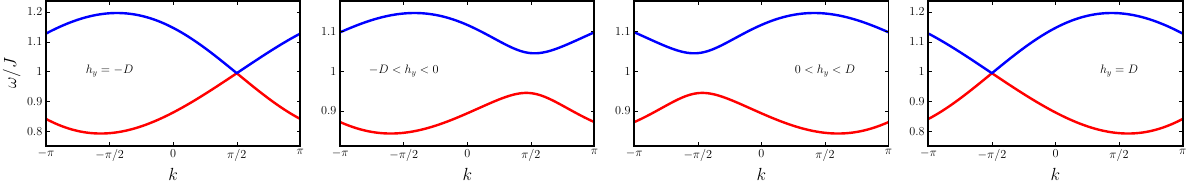}
\caption{Triplon bands ($t_{x}$ and $t_{z}$) obtained from $\mathcal{H}_{k}$, Eq.~\eqref{triplon_ham_k}, are plotted for different $h_y$. 
We see that the gap between the two modes vanishes to form a Dirac point at  
$\hy=\pm\dm$. Everywhere else in the dimer-quantum-paramagnetic phase, the two modes do not touch each other. 
For $|h_y|<D$ the phase is topologically non-trivial, else it is trivial. The parameters used  are $\dm/\jp=\gm/\jp=0.1$ and 
$\jn/\jp=0.01$.}
\label{fig:modes}
\end{figure*}
%%%%%
%\end{widetext}

\begin{figure}
\centering
\includegraphics[width=0.4\textwidth]{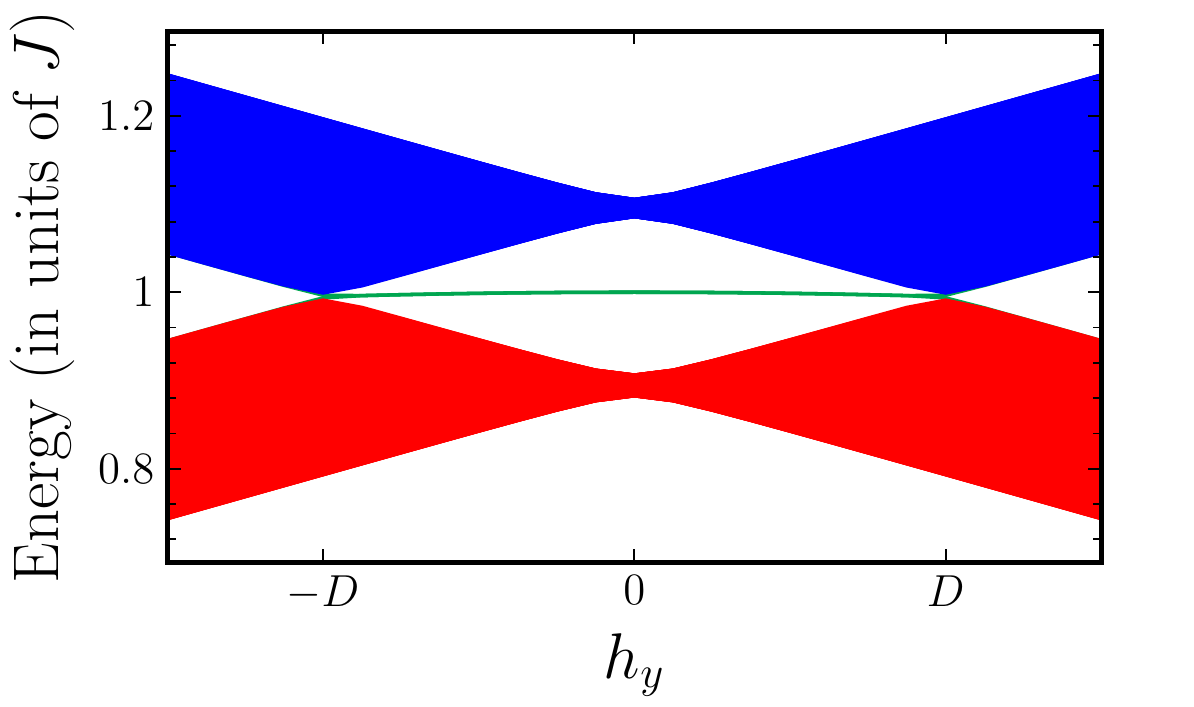}
\caption{Band structure of the quantum spin ladder, Eq.~\eqref{triplon_ham_k}, with open ends. Protected end states  (green line) appear in the topological paramagnetic phase, $|h_y|<D$. Parameters used are the same as in Fig.~\ref{fig:modes}.
%{\clr [are they doubly degenerate at each end?-No]}
}
\label{fig:es}
\end{figure}

\begin{figure}[t]
\centering
\includegraphics[width=0.4\textwidth]{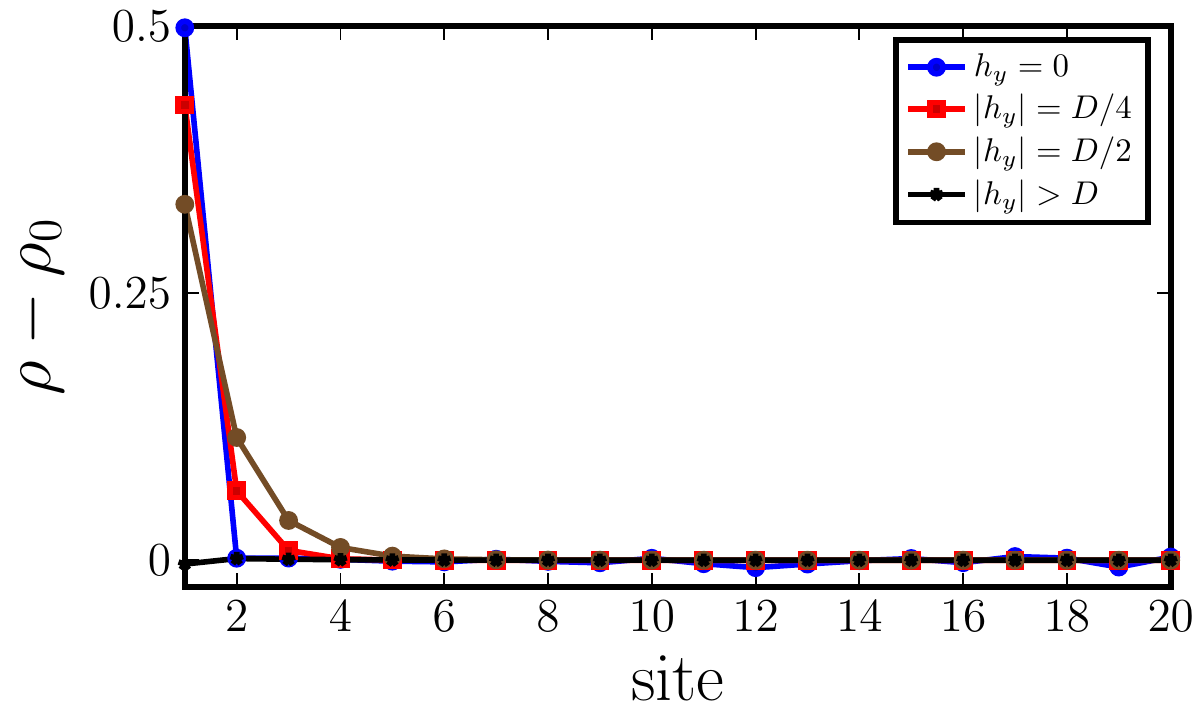} 
\caption{
Triplon end-state density profile $\rho- \rho_0$
 plotted in the topological paramagnetic phase, $|h_y|<D$, near one of the ladder ends. 
In the topologically trivial phase, $|h_y| > D$, the end states are absent (black trace). 
%The inset shows the particle quantum number ($\mathcal{N}_{t}$) of triplon end states. 
Parameters used are the same as in Fig.~\ref{fig:modes}.}
\label{fig:ldos}
\end{figure}

The elementary low-energy excitations of  coupled-dimer magnets correspond to breaking a singlet dimer
 into a spin-1 triplet state. These excitations are called {\em triplons} and can be viewed as
 bosonic quasiparticles with S=1. In the absence of spin-orbit coupling the three triplet states 
 are degenerate, due to SU(2) spin-rotation symmetry. For spin-ladder compounds with heavy elements,
 however, strong spin-orbit interactions lead to   antisymmetric DM  couplings, 
 %and a symmetric interaction between different spin components of neighbouring  spins.
 %Besides the DM interaction, SOC also produces a symmetric interaction between different spin component of neighboring spins along the legs
 which split the triplon band into multiple dispersive bands.
 We find that these triplon bands  can have  nontrivial topological character,
 which can be tuned by an applied field. In the  topologically nontrival phase, which we call  
 \emph{topological quantum paramagnet}, the spin-ladder exhibits triplon end-states with fractional
 particle number (see Figs.~\ref{fig:es} and \ref{fig:ldos}).  We show that  these end-states are protected by a nonzero winding number 
 and determine their experimental signatures in heat-transport and neutron-scattering measurements. 
 
%spin-orbit coupling which introduces frustration. 
%with frustration arising from spin orbit coupling. 

\emph{Spin model and triplon description.--} 
We consider  a spin-$1/2$ frustrated quantum spin ladder, whose lattice geometry and interactions are illustrated in Fig.~\ref{fig:model}.
The corresponding Hamiltonian is given by
\begin{eqnarray}
\mathcal{H} 
&&= 
\jp \sum_{i} \vec{S}_{1i} \cdot \vec{S}_{2i} 
+ \jn \sum_{i} \big[ \vec{S}_{1i} \cdot \vec{S}_{1i+1} + \vec{S}_{2i} \cdot \vec{S}_{2i+1} \big] \nonumber \\
&&+ \dm \sum_{i} \big[ S^{z}_{1i} S^{x}_{1i+1} - S^{x}_{1i} S^{z}_{1i+1} 
+ S^{z}_{2i} S^{x}_{2i+1} - S^{x}_{2i} S^{z}_{2i+1} \big] \nonumber \\
&&+ \gm \sum_{i} \big[ S^{z}_{1i} S^{x}_{1i+1} + S^{x}_{1i} S^{z}_{1i+1} 
+ S^{z}_{2i} S^{x}_{2i+1} + S^{x}_{2i} S^{z}_{2i+1} \big] \nonumber \\
&&+ \hy \sum_{i} \big[ S^{y}_{1i} + S^{y}_{2i} \big] ,
\label{eq:model}
\end{eqnarray}
where, $i$ denotes the dimer site, $1,2$ label the two legs of the ladder, $\jp$ is the antiferromagnetic intra-dimer coupling, and $\jn$ is the inter-dimer Heisenberg interaction. Spin-orbit coupling gives rise to the odd-parity DM interaction $\dm$ and the even-parity spin-anisotropic inter-dimer coupling $\gm$.
We assume that the two legs of the ladder are equivalent by symmetry. Likewise, all the rungs are taken to be equivalent. 
Therefore, the only symmetry allowed DM term is the inter-dimer DM interaction in the $y$-direction between the spins along the legs of the ladder~\cite{note}.
 The even-parity spin-anisotropic interaction $\gm$ is of similar form as the DM term, but its direction is not fixed by lattice symmetries. For simplicity, we
assume that the $\gm$ term points in the same direction as the DM interaction; in the Supplemental Material we consider the
case where the $\gm$ term points along the $z$ direction.
In Eq.~\eqref{eq:model} we have also included a small magnetic field $\hy$ perpendicular to the ladder plane, which provides
a handle to induce a topological transition.

For dominant   $\jp>0$, the spins within each unit cell of the spin ladder form a singlet and a dimer-quantum-paramagnet is realized. Throughout this paper we shall be interested in this phase only. This phase has three gapped excitations corresponding to the three possible spin-$1$ triplet excited states on each dimer.
To describe these elementary triplon excitations, we employ the bond-operator formalism~\cite{ss_bhatt}, which allows us to represent
the spin operators in Eq.~\eqref{eq:model} in terms of triplon creation and annihilation operators $t^{\dag}_\gamma$ and $t^{\ }_\gamma$ ($\gamma=x,y,z$)~\cite{supp}.
For a given dimer these triplon operators are defined as $t_{\gamma}^{\dagger} |t_0\rangle = |t_\gamma \rangle$ ($\gamma=x,y,z$), where $|t_0\rangle = [|\uparrow\downarrow\rangle - |\downarrow\uparrow\rangle]/\sqrt{2}$ is the singlet state, while $|t_x\rangle = -[|\uparrow\uparrow\rangle - |\downarrow\downarrow\rangle]/\sqrt{2}$, $|t_y\rangle = \ii[|\uparrow\uparrow\rangle + |\downarrow\downarrow\rangle]/\sqrt{2}$, and 
$|t_z\rangle = [|\uparrow\downarrow\rangle + |\downarrow\uparrow\rangle]/\sqrt{2}$ are the  spin-$1$ triplet states. 
Rewriting Eq.~\eqref{eq:model} in terms of $t_\gamma$ and $t^{\dag}_\gamma$ yields an interacting bosonic Hamiltonian describing
the dynamics of the triplons~\cite{supp}. 
For simplicity, we consider here only the bilinear part of this triplon Hamiltonian.
This is known as the {\em harmonic approximation} \cite{ha_note}. 
%and has been shown to be a controlled approximation in large dimensions \cite{larged_para, larged_af}. 
As it turns out, at the harmonic level the $t_{y}$ triplon mode is decoupled from the other two triplons.
We therefore focus only on the $t_{x}$ and $t_{z}$ excitations, whose dynamics in momentum space
is given by~\cite{supp}
%for the $t_y$ mode; it reads as $\omega_{k y} = \sqrt{(\jp+\jn \cos(k))^2 - \jn^2}$.  
%
%Using periodic boundary condition and lattice translation symmetry, we can introduce the corresponding triplon operators in the 
%Fourier space and thus obtain the following form for the bilinear Hamiltonain involving $t_{x}$ and $t_{z}$ modes in the momentum 
%space:
\begin{subequations} \label{triplon_ham_k}
\begin{equation}
\mathcal{H}_{k} = \frac{1}{2} \sum_{k} \Psi^{\dagger}_{k} \mathcal{M}_{k} \Psi_{k} ,
\label{eq:ham_k}
\end{equation}
with the spinor
$
\label{eq:psi}
\Psi_{k} = 
(
t_{kx} , t_{kz} ,  t^{\dagger}_{-kx} ,  t^{\dagger}_{-kz}
)^{T} $
%%%%%%
and the $4 \times 4$ matrix
\begin{align}
\mathcal{M}_{k} &= 
\begin{bmatrix}
H_{1}(k) & H_{2} (k) \\
H^{\dagger}_{2} (k) & H^{T}_{1} (-k) 
\end{bmatrix} \,.
\end{align}
\end{subequations}
The diagonal and off-diagonal parts of $\mathcal{M}_k$ read
\begin{subequations}
\begin{align}
\label{eq:h1}
H_{1} (k) &= 
(\jp + \jn \cos (k)) \mathbbm{1} + \vec{d} \cdot \vec{\sigma} ,\\
%%%%%%%%
\label{eq:h2}
H_{2} (k) &= 
-\jn e^{-\ii k} \mathbbm{1} - \vec{x} \cdot \vec{\sigma} \, ,
\end{align}
with the vectors
\begin{align}
\label{eq:dk}
\vec{d} &\equiv \left\lbrace d_{1}, d_{2}, d_{3} \right\rbrace = \left\lbrace \gm \cos (k), -\dm \sin (k) -\hy, 0 \right\rbrace \,, \\
%%%%
\label{eq:xk}
\vec{x} &\equiv \left\lbrace x_1, x_2, x_3 \right\rbrace = \left\lbrace \gm \cos(k), -\dm \sin(k), 0 \right\rbrace \,,
\end{align}
\end{subequations}
where
$\mathbbm{1}$ is the $2 \times 2$ identity matrix, $\ii=\sqrt{-1}$ and $\vec{\sigma} \equiv \left\lbrace \sigma_1, \sigma_2, \sigma_3 \right\rbrace$ are 
the three Pauli matrices. 

%%%%%
%%%%%

\emph{Triplon bands and protected end states.--}
The triplon bands of Hamiltonian~\eqref{triplon_ham_k} are obtained  by use of a bosonic Bogoliubov transformation \cite{diab, dia1, dia2}, which amounts to 
diagonalizing the non-Hermitian matrix $\Sigma \mathcal{M}_{k}$, where $\Sigma = \textrm{diag}(\mathbbm{1}, -\mathbbm{1})$.
In Fig.~\ref{fig:modes} we show
the typical triplon dispersions for different values of the tuning parameter $\hy$. 
Both triplon modes are gapped in the entire dimer-quantum-paramagnetic phase. Moreover, the two triplons do not touch each other, except at 
$\hy=\pm\dm$, where they touch linearly. 
This observation suggest that at $\hy=\pm\dm$ 
there occurs a topological phase transition, which separates a trivial phase 
from a topological one. 

To confirm this conjecture, we study the edge  states of Hamiltonian $\mathcal{H}_{k}$, whose presence
indicates the topological character of the triplon bands. 
For that purpose we determine the eigenenergies and eigenmodes of $\mathcal{H}_{k}$  in real space with open boundary conditions.
Figure~\ref{fig:es} displays the so-obtained spectrum as a function of $\hy$. 
We also compute the energy-integrated local density of states  (LDOS) of $\mathcal{H}_{k}$, by adding the contributions from 
the lower  triplon band and from the end states with energies in between the two triplon bands.
%(i.e., all states with energy $E \leq 1$).
To reveal the existence of end states we subtract  the LDOS of $\mathcal{H}_k$ with periodic boundary conditions $\rho_0$
from the LDOS with open boundary conditions $\rho$.
The resulting triplon end-state density profile $\rho- \rho_0$ is plotted 
in Fig.~\ref{fig:ldos} for different values of  $\hy$. 
From Fig.~\ref{fig:es} we clearly see that   for $|\hy|<\dm$ the
spectrum contains, besides the bulk triplon bands (red and blue), an additional state (green)
with energy in between the two triplons.   
Figure~\ref{fig:ldos} shows that this in-gap state is exponentially localized at the two ends of the spin ladder. 
Hence, we conclude that the  paramagnetic phase of S=1/2 quantum spin ladders  
is subdivided into a trivial phase ($|\hy|>\dm$) and a topological phase ($|\hy|<\dm$) 
\cite{footnote_gs}. 
We call the latter a {\em topological quantum paramagnet}~\cite{footnote_QSL}, which is characterized by
a non-zero winding number, as we will show below.

%{\clr
But before doing so, let us examine the area
under the peaks in the triplon end-state density profile  of Fig.~\ref{fig:ldos}.
We find that it is zero in the trivial phase, while
in the topological phase it takes on the fractional value $1/2$. 
This fractional value is reminiscent of the charge $e/2$ end states in the Su-Schrieffer-Heeger model~\cite{ssh_review} 
and is intimately connected to the nontrivial topology of the system~\cite{qi_zhang}. Physically, the fractional value hints towards a fractionalized nature of the triplon end states. However, unlike the SSH model, here we are dealing with bosons and it is not straightforward to establish this connection. This will be addressed in future work.
%}

%%%%%
%%%%%

\emph{Winding number.--}
We now show that the topological quantum paramagnetic phase is characterized by a non-zero winding number. Although the problem at hand 
is seemingly similar to a one-dimensional fermionic topological insulator, we find that  
the calculation of the winding number proceeds along quite different lines than in the fermionic case. 
Recall that in order to compute the winding number of fermionic systems, one first needs to identify the chiral symmetry operator and transform
 the Hamiltonian to a basis 
wherein the chiral symmetry operator is diagonal. This results in a   block off-diagonal Hamiltonian, which is then used to calculate the winding 
number~\cite{classif2,classif1}. 
For our bosonic model, we find that   Eq.~\eqref{triplon_ham_k} can be deformed into 
a chiral symmetric Hamiltonian, i.e., for $\jn=0$ we have 
$\left\lbrace \mathbbm{1} \otimes \sigma_3 , \mathcal{M}_{k}- \jp \mathbbm{1} \otimes \mathbbm{1} \right\rbrace = 0$, 
since  $\sigma_3$ anticommutes with $H_1 - (\jp+\jn\cos(k)) \mathbbm{1}$ and with
$H_2 + \jn e^{-\ii k} \mathbbm{1}$.   
However, this observation is not very helpful for two reasons: (i) the symmetry operator is already diagonal and (ii) the eigenmodes of our 
model are not given by $\mathcal{M}_{k}$, but rather by $\Sigma \mathcal{M}_{k}$.

%%%%%%
%\begin{figure}
%\centering
%\includegraphics[width=0.4\textwidth]{./images/plot_charge_01}
%\caption{
%Particle number of the triplon end state as a function of applied field $h_y$.
%In the  topological paramagnetic phase the particle number is $1/2$, while it is zero in the  trivial paramagnetic phase.
%Parameters used are the same as in Fig.~\ref{fig:modes}. 
%{\clr [Maybe better not to call it charge? Is the small dip at $h_y=0$ due to numerics?]}}
%\label{fig:charge}
%\end{figure}
%%%%%%

\begin{figure}[t]
\centering
\includegraphics[width=0.4\textwidth]{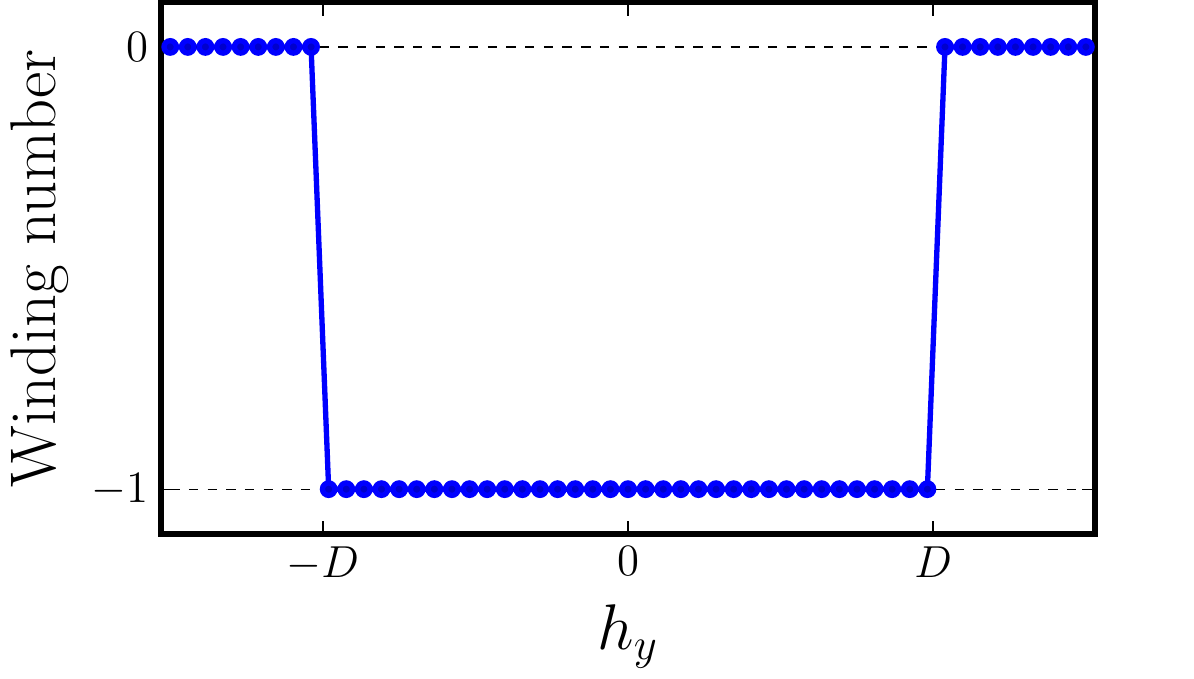}
\caption{Winding number $\mathcal{W}$, Eq.~\eqref{eq:win}, as a function of applied field $h_y$. 
In the topological paramagnetic phase, $|h_y|<D$,  $\mathcal{W}$ evaluates to $-1$, which, 
by the bulk-boundary correspondence, leads to the appearance of triplon end states, cf.\ Figs.~\ref{fig:es} and~\ref{fig:ldos}.
Parameters used are the same as in Fig.~\ref{fig:modes}.}
\label{fig:win}
\end{figure}

Hence, we need to find another way to bring  $\Sigma \mathcal{M}_{k}$ into block off-diagonal form. 
To that end, let us consider the   transformation with the unitary matrix 
\begin{equation}
\label{eq:ut}
U = 
\begin{bmatrix}
0 & 0 & 1 & 0 \\
1 & 0 & 0 & 0 \\
0 & 0 & 0 & 1 \\
0 & 1 & 0 & 0 
\end{bmatrix} \,.
\end{equation}
Under the action of $U$, the relevant matrix $\Sigma \mathcal{M}_{k}$ transforms as 
\begin{subequations}
\begin{align}
\label{eq:qq}
\tilde{\mathcal{M}_{k}} &= U^{\dagger} \left(\Sigma \mathcal{M}_{k} \right) U  
= \begin{bmatrix}
\mathcal{A}_{k} & \mathcal{D}_{1 k} \\
\mathcal{D}_{2 k} & \mathcal{A}_{k} 
\end{bmatrix} \,,
\end{align}
where
\begin{align}
\label{eq:ak}
\mathcal{A}_{k} &= \begin{bmatrix}
\jp + \jn \cos(k) & -\jn e^{-\ii k} \\  
\jn e^{\ii k} & -\jp - \jn \cos(k)
\end{bmatrix} \,,  
\end{align}
and the off-diagonal blocks are given by
%%%%%
\begin{align}
\label{eq:d1}
\mathcal{D}_{1 k} &=
\begin{bmatrix} 
x_1 + \ii (x_2 - \hy) & -x_1 - \ii x_2 \\
x_1 + \ii x_2  & -x_1 - \ii (x_2 + \hy) 
\end{bmatrix} \,, \\
%%%%%
\label{eq:d2}
\mathcal{D}_{2 k} &=
\begin{bmatrix}
x_1 - \ii (x_2 - \hy) & -x_1 + \ii x_2  \\
x_1 - \ii x_2  & -x_1 + \ii (x_2 + \hy)  
\end{bmatrix} \,.
\end{align}
\end{subequations}
Although $\tilde{\mathcal{M}_{k}}$ is not  block off-diagonal, note that the diagonal block $\mathcal{A}_{k}$ only leads to an overall energy shift 
(same for both modes) and small variations in the shape of the modes, 
but does not alter the topological properties.
%However, $\mathcal{A}_{k}$ does not influence the Dirac point Physics or the topological aspect. 
This is most easily seen by noting that the difference in the triplon energy spectrum with or without the anomalous terms $\mathcal{H}_{2} (k)$
%{\clr [shouldn't this be $\mathcal{A}_{k}$?]}
 is negligible. So let us focus on $\mathcal{D}_{1 k}$ and $\mathcal{D}_{2 k}$. In a 
way similar to the fermionic case, we can define the winding number as
\begin{equation}
\label{eq:win}
\mathcal{W} = \frac{1}{2} \frac{1}{4 \pi \ii} \int_{BZ} dk ~ Tr \big[ \mathcal{D}^{-1} \partial_{k} \mathcal{D} 
- (\mathcal{D}^{\dagger})^{-1} \partial_{k} \mathcal{D}^{\dagger} \big] \,,
\end{equation}
where $\mathcal{D} =  ( \mathcal{D}_{1 k} + \mathcal{D}^{\dagger}_{2 k} ) / 2 $.
We note that the factor $1/2$ in Eq.~\eqref{eq:win} is due to the prefactor $1/2$ in Eq.~\eqref{eq:ham_k}. 
The winding number $\mathcal{W}$ is quantized to integer values and evaluates to 
$\mathcal{W} = -1$ in the topological quantum paramagnetic phase $|h_y|<D$, see Fig.~\ref{fig:win}. 
By the bulk-boundary correspondence, the non-zero winding number leads to the protection of the triplon end-states of Fig.~\ref{fig:modes}.

%%

%%%%%
%%%%%

\emph{Conclusions and implications for experiments.--}
We have studied  topological properties of   
 S=1/2 quantum spin ladders with strong spin-orbit coupling and
  have shown that the quantum-disordered paramagnetic state
of these spin ladders subdivides into a  trivial and a  topological phase.
The latter is, what we call, a topological quantum paramagnet, since it  
exhibits topologically nontrivial triplon excitations.
It should be noted that there is no qualitative difference between the ground states in the two phases. 
The topological aspects feature only in the triplon excitation modes. 
The phase transition between the topological and the trivial quantum paramagnet
can be tuned by an applied field and
occurs when two triplon modes touch, forming a Dirac point. 
The topological quantum paramagnet  has a non-trivial winding number,
which leads to 
protected triplon end states with fractional particle number $1/2$.
%A more complex situation arises when all the three triplon modes are coupled, which is discussed in~\cite{supp}. {\clr [?] }

We expect that the topological quantum paramagnetic phase exists in many spin ladder compounds, even for relatively weak 
spin-orbit interactions~\cite{footnote_weak_soc}.
The quantum dimer model of Eq.~\eqref{eq:model} is just one example   
 of a large class of Hamiltonians that all exhibit the same topological phase. It is always possible to add small perturbations to 
 Hamiltonian~\eqref{eq:model} without changing its topological properties. Based on these considerations we expect that topological triplon bands
are quite ubiquitous.
 A particularly promising candidate material 
 for observing topological triplons
 is the strongly spin-orbit coupled spin ladder BiCu$_2$PO$_6$, for which field induced phase transitions have
recently been investigated \cite{bcp_exp, bcp_th}. 
To experimentally study the topological phase transition and the evolution of  the triplon band structure as a function of applied field one may use neutron scattering experiments.
It may even be possible to directly observe the triplon end states using small-angle neutron scattering (SANS). In fact, 
our calculations show that 
the local dynamic spin structure factor exhibits a sharp peak at the triplon end-state energy (see~\cite{supp}), which should be observable in SANS. %{\clr [Add comment about damping]}
Another possibility is to use specific heat measurements to look for the residual $\ln 2$ entropy contributed by the triplon end states. 
%Since the end states have a fractional particle number of $1/2$, this corresponds to a residual  entropy of $\ln 2$. 

The triplon interaction terms, arising beyond the harmonic approximation, can in principle result in the intrinsic zero-temperature damping of the triplon modes \cite{decay_review}.
%{\clb 
This could in particular also apply %} 
to the localized end states in the topological phase \cite{topo_decay}. However, as long as the gap is greater than $0.5\jp$, because of energy-momentum constraints, the end states will not decay spontaneously. %{\clb 
This is in contrast to the topological edge excitations of
ordered magnets~\cite{weyl_magnon,sk_kim}, which can decay by coupling to the Goldstone modes. %}

Our findings represent the first step towards the development of a comprehensive topological band theory for triplons.  Indeed, we expect  
the topological quantum paramagnet to be a rather commonly occurring phase, which may exist even in two-  and
three-dimensional quantum magnets. One possible generalization of our work are two-dimensional magnets composed of coupled  spin-ladders, which may exhibit dispersing triplon edge modes carrying dissipationless spin current.
Besides this, other interesting questions for further study are: (i) the fate of the topological quantum paramagnet at finite temperature and (ii) the study of phase transitions between   topological quantum paramagnets and   quantum spin liquids.

% \emph{Acknowledgments.---}
We thank A.~Chernyshev, G.~Khaliullin, S.~Rachel, H.~Takagi, and M.~Vojta for useful discussions. A.P.S. is grateful to the KITP at UC Santa Barbara for hospitality during the preparation of this work. This research was supported in part by the National Science Foundation under Grant No. NSF PHY-1125915.

 \bibliographystyle{apsrev}
\bibliography{bib_topo_para_arxiv_v3}

\begin{thebibliography}{48}
\bibitem[{ha_()}]{ha_note}
\bibinfo{note}{Inside the paramagnetic phase and away from a magnetic quantum
  critical point, the triplon density is small and hence triplon interactions
  can be neglected. In general, the harmonic approximation is controlled in
  large dimensions, both in the paramagnetic as well as inside the magnetically
  ordered phases, as shown in Refs.~\cite{larged_para,larged_af}. That said,
  small interactions cannot change the topology of the quantum spin ladder,
  since they respect the same symmetries of the spin Hamiltonian, which are
  given by the lattice.}
\bibitem[{foo({\natexlab{a}})}]{footnote_gs}
\bibinfo{note}{Note that there is no qualitative difference between the ground
  states of the topologically trivial and topologically non-trivial phases. In
  both the phases, there are triplon excitations over the singlet product
  state. The topological aspect, which distinguishes the two phases, features
  in these triplon bands and is amenable to experimental detection as discussed
  in the text. However, unlike in fermionic topological systems (e.g.,
  topological superconductors), these experimental signatures are observed at
  an elevated energy rather than at zero energy.}
\bibitem[{foo({\natexlab{b}})}]{footnote_QSL}
\bibinfo{note}{Note that the topological quantum paramagnet studied in this
  Letter is different from a quantum spin liquid, which is characterized by
  fractional bulk excitations and topological order. Three-dimensional quantum
  spin liquids with fractional bulk excitations have recently been studied in
  Refs.~\cite{av_ts,cw_ts} using field theory methods.}
\bibitem[{foo({\natexlab{c}})}]{footnote_weak_soc}
\bibinfo{note}{The topological phase occurs whenever the DM interaction $D$ is
  larger than the applied magnetic field $h_y$. It should be noted, however,
  that the localized end states are most clearly visible only in the regime $K
  < D$, where there exists a full gap between the two triplon bands. In
  contrast, in the regime $K > D$ the triplon end states overlap with the bulk
  states and are therefore broadened and decay algebraically into the bulk.}
\end{thebibliography}

%%%%%%%%%%%%%%%%%%%%%%%%%%%%%%%%%%%%%%%%%%%%%%%%%%
%%% Supp. Mat.
%%%%%%%%%%%%%%%%%%%%%%%%%%%%%%%%%%%%%%%%%%%%%%%%%%

%%%%%%%%%%%%%%%%%%%%%%% Supplemental material %%%%%%%%%%%%%%%%%%%%%%%%%%%%%
\clearpage

%%%%%%%%%%
\setcounter{equation}{0}
\setcounter{figure}{0}
\setcounter{table}{0}
\setcounter{page}{1}
\renewcommand{\theequation}{S\arabic{equation}}
\renewcommand{\thefigure}{S\arabic{figure}}
\renewcommand{\bibnumfmt}[1]{[S#1]}
\renewcommand{\citenumfont}[1]{S#1}
%%%%%%%%%%

%\title[]
%{ Supplemental material: \\
%Topological quantum paramagnet in a quantum spin ladder
%}
%\author{Darshan G. Joshi}
%\author{Andreas P. Schnyder}
%\affiliation{Max Planck Institute for Solid State Research, D-70569 Stuttgart, Germany}
%
%\date{\today}
%
%\maketitle

%\begin{widetext}

%\onecolumn
\onecolumngrid
\begin{center}
\textbf{\large Supplemental material: \\
Topological quantum paramagnet in a quantum spin ladder} \\
\medskip 

Darshan G. Joshi and Andreas P. Schnyder \\
\smallskip

{\em Max-Planck-Institute for Solid State Research, D-70569 Stuttgart, Germany}

\end{center}
%\end{widetext}

\bigskip

\twocolumngrid
%%%%%%%%%%%%%%%%%%%%%%%%%%%%%%%%%%%
%%%%%%%%%%%%%%%%%%%%%%%%%%%%%%%%%%%
\section{Triplon Hamiltonian}

As discussed in the main text, the spin-1 excitations on top of a singlet state are described as bosonic quasiparticles, triplons, 
within the bond-operator theory~\cite{ss_bhatt_s}. The spin operators can be thus expressed as follows in terms of triplon operators:
\begin{align}
\label{eq:s_op}
S^{\alpha}_{1,2i} &= \frac{1}{2} \big[ \pm \ii t^{\dagger}_{i\alpha} P_{i} \mp \ii P_{i} t_{i\alpha} 
- \ii \epsilon_{\alpha \beta \gamma} t^{\dagger}_{i\beta} t_{i\gamma} \big] \,, \\
%%%%
\label{eq:s1s2}
\vec{S}_{1i} \cdot \vec{S}_{2i} &= -\frac{3}{4} + \sum_{\alpha} t^{\dagger}_{i\alpha} t_{i\alpha} \,.
\end{align}
Here, $t_{\alpha}$ ($\alpha=x,y,z$) are the triplon operators and $P_{i}=1-\sum_{\alpha} t^{\dagger}_{i\alpha} t_{i\alpha}$ is the 
projection operator \cite{collins_s} which takes care of the Hilbert-space constraint, i.e., no more than one boson per dimer. Note that 
the appearance of $\ii=\sqrt{-1}$ in the above equation is due the convention in which the triplon operators are time-reversal 
invariant \cite{tr_bop_s}. Inserting this in the spin Hamiltonian [Eq. (1)] in the main text we obtain the triplon Hamiltonian as follows:
\begin{align}
\mathcal{H} &= -\frac{3}{4} \jp N + \jp \sum_{i \alpha} t^{\dagger}_{i\alpha} t_{i\alpha}  \nonumber \\
&+ \frac{\jn}{2} \sum_{i\alpha} \big[ t^{\dagger}_{i\alpha} P_{i} P_{i+1} t_{i+1 \alpha}
- t^{\dagger}_{i\alpha} P_{i} t^{\dagger}_{i+1\alpha} P_{i+1} + H.c. \big] \nonumber \\
&+\frac{\dm}{2} \sum_{i} \big[ t^{\dagger}_{iz} P_{i} P_{i+1} t_{i+1 x} - t^{\dagger}_{iz} P_{i} t^{\dagger}_{i+1x} P_{i+1}  
\nonumber \\
&~~~~~~~~~~- t^{\dagger}_{ix} P_{i} P_{i+1} t_{i+1 z} + t^{\dagger}_{ix} P_{i} t^{\dagger}_{i+1z} P_{i+1} + H.c. \big] \nonumber \\
&+\frac{\gm}{2} \sum_{i} \big[ t^{\dagger}_{iz} P_{i} P_{i+1} t_{i+1 x} - t^{\dagger}_{iz} P_{i} t^{\dagger}_{i+1x} P_{i+1} 
\nonumber \\
&~~~~~~~~~~+ t^{\dagger}_{ix} P_{i} P_{i+1} t_{i+1 z} - t^{\dagger}_{ix} P_{i} t^{\dagger}_{i+1z} P_{i+1} + H.c. \big] \nonumber \\
&+\ii \hy \sum_{i} \big[ t^{\dagger}_{ix} t_{iz} - t^{\dagger}_{iz} t_{ix} \big] 
\label{eq:ham_r}
\end{align}
where, $N$ is number of dimer sites. Here on, we shall consider only the bilinear piece in the above interacting Hamiltonian. This is 
known as the {\em harmonic approximation} and it has been shown that it is a controlled approximation, such that corrections beyond it can be arranged in a systematic expansion in inverse spatial dimension \cite{larged_para_s, larged_af_s}. So triplon interactions 
will not qualitatively change the physics discussed here. The bilinear Hamiltonian is as follows:
\begin{align}
\mathcal{H}_{2} &= \jp \sum_{i \alpha} t^{\dagger}_{i\alpha} t_{i\alpha}  %\nonumber \\
+ \frac{\jn}{2} \sum_{i\alpha} \big[ t^{\dagger}_{i\alpha} t_{i+1 \alpha}
- t^{\dagger}_{i\alpha} t^{\dagger}_{i+1\alpha}  + H.c. \big] \nonumber \\
&+\frac{\dm}{2} \sum_{i} \big[ t^{\dagger}_{iz} t_{i+1 x} - t^{\dagger}_{iz} t^{\dagger}_{i+1x}  
- t^{\dagger}_{ix} t_{i+1 z} + t^{\dagger}_{ix} t^{\dagger}_{i+1z}  + H.c. \big] \nonumber \\
&+\frac{\gm}{2} \sum_{i} \big[ t^{\dagger}_{iz} t_{i+1 x} - t^{\dagger}_{iz} t^{\dagger}_{i+1x}  
+ t^{\dagger}_{ix} t_{i+1 z} - t^{\dagger}_{ix} t^{\dagger}_{i+1z} + H.c. \big] \nonumber \\
&+\ii \hy \sum_{i} \big[ t^{\dagger}_{ix} t_{iz} - t^{\dagger}_{iz} t_{ix} \big] 
\label{eq:ham_har}
\end{align}

At the harmonic level, the $t_{y}$ mode is decoupled from the other two triplon modes. It is straightforward to obtain its 
dispersion; $\omega_{k y} = \sqrt{(\jp+\jn \cos(k))^2 - \jn^2}$. We shall now on focus only on the $t_{x}$ and 
$t_{z}$ triplon modes, which is discussed in the main text. 

%%%%%%%%%%%%%
\begin{figure*}
\includegraphics[width=1.0\textwidth]{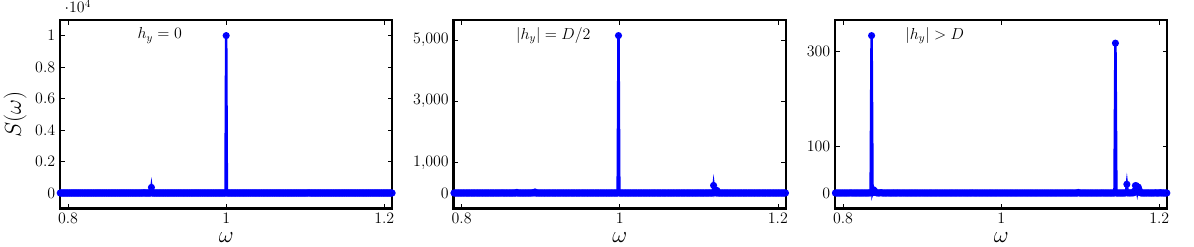}
\caption{Dynamic structure factor at $k=0$ for different values of external field $\hy$. We see sharp peaks at the end-state energy 
within the topological-quantum-paramagnetic phase, while these peaks disappear in the trival paramagnetic phase.
The parameters used here are $\dm/\jp=\gm/\jp=0.1$ and $\jn/\jp=0.01$.}
\label{fig:s_w}
\end{figure*}
%%%%%%%%%%%%%

%%%%%%%%%%%%%%%%%%%%%%%%%%%%%%%%%%%

%%%%%%%%%%%%%%%%%%%%%%%%%%%%%%%%%%%
\section{Local density of states}

In the main text, we presented the local density of states to establish that the end states were indeed localized at the ends of the 
ladder. Here we sketch the necessary technical details. As mentioned in the main text, the eigenmodes of the triplon Hamiltonian are 
obtained by diagonalizing the matrix $\Sigma \mathcal{H}$. This is a non-Hermitian matrix and it is diagonalized in the following 
way \cite{diab_s, dia1_s, dia2_s}:
\begin{equation}
\label{eq:diag_ham}
\Omega=T^{\dagger} \mathcal{H} T
\end{equation}
where, $\Omega$ is the diagonal matrix containing the eigenmodes of the Hamiltonian and 
\begin{equation}
\label{eq:t_def}
T=
\begin{bmatrix}
U & V \\
V^{*} & U^{*} 
\end{bmatrix}
\end{equation}
satisfying the condition,
\begin{equation}
T^{\dagger} \Sigma T = T \Sigma T^{\dagger} = \Sigma \,.
\end{equation}
This ensures that the bosonic commutation relations are satisfied. 
The matrices $U$ and $V$ contain the Bogoliubov coefficients. For an eigenfrequency $\omega_n$,
\begin{equation}
\label{eq:uv}
\Sigma \mathcal{H} |\phi_n \rangle = \omega_n |\phi_n \rangle \,; ~~~~
|\phi_n \rangle = 
\begin{bmatrix}
u_n \\
v^{*}_{n}
\end{bmatrix} \,.
\end{equation}
The local density of states as a function of site i is then given by
\begin{equation}
\label{eq:dos}
\rho(i) = \sum_{n \in \lbrace \omega_n \leq \omega_{0} \rbrace} 
\bigg[ |u_{i,n}|^{2} + |u_{i+N,n}|^{2} - |v_{i,n}|^{2} - |v_{i+N,n}|^{2} \bigg] 
\end{equation}
where, $\omega_{0}$ is the end-state energy. This is plotted in Fig.~4 in the main text, where we see exponentially localized peaks near the ladder ends inside the topological phase.

%%%%%%%%%%%%%%%%%%%%%%%%%%%%%%%%%%%

%%%%%%%%%%%%%%%%%%%%%%%%%%%%%%%%%%%
\section{Winding number}

In case of a hopping Hamitonian in the Dirac form, it is straightforward to calculate the winding number invariant. However, for the model discussed in the main text, we also have anomalous terms. As mentioned in the main text, unlike the fermionic case, it is not straightforward to bring the Hamiltonian in an off-block diagonal form. With the help of a suitable transformation, however, we were able to write down an expression for the winding number [Eq. (6)]. Here we quote the explicit expression obtained after inserting the matrices $\mathcal{D}_{1k}$ and $\mathcal{D}_{2k}$ in Eq.~(6) in the main text:
\begin{widetext}
\begin{equation}
\label{eq:win_for}
\mathcal{W}=\frac{1}{8\pi} \int_{-\pi}^{\pi}  \frac{2 \dm \gm (-\dm^2 - \gm^2 + (\dm^2 - \gm^2 - 2 \hy^2) \cos(2k))}{
\gm^4 \cos^{4}(k) + (\hy^2 - \dm^2 \sin^{2}(k))^2 + 2\gm^2 \cos^{2}(k) (\hy^2 + \dm^2 \sin^{2}(k))}  
\end{equation}
\end{widetext}
This is plotted in Fig. (5) in the main text. As is evident, the winding number is non zero in the topological quantum paramagnetic phase, while zero in the trivial paramagnetic phase. 

Note that although both $\dm$ and $\gm$ enter the formula for the winding number, the topological phase transition depends only on the relative values of $\hy$ and $\dm$. Any non-zero value of $\gm$ is sufficient to keep the topological aspect. For $\gm=0$ or $\dm=0$, the winding is always zero and hence there will be no topological quantum paramagnet.

%%%%%%%%%%%%%%%%%%%%%%%%%%%%%%%%%%%

%%%%%%%%%%%%%%%%%%%%%%%%%%%%%%%%%%%
\section{Dynamic structure factor}

We wish to calculate the dynamic structure 
factor for the ladder with open boundaries to see the signature of the end states. But this means that the crystal momentum is not a good quantum number anymore. So we calculate the {\em local} dynamic structure factor, which will correspond to the $k=0$ contribution 
in a scattering experiment. The dynamic structure factor,
\begin{equation}
\label{eq:sf_def}
S(k,\omega) = \frac{1}{N} \sum_{i,j} S_{ij} e^{\ii k r_{ij}} \,.
\end{equation}
So,
\begin{equation}
\label{eq:s_w}
S(\omega) \equiv S(k=0,\omega) = \frac{1}{N} \sum_{i,j} S_{ij} \,,
\end{equation}
where $S_{ij} = - \Im \chi_{ij}$, with $\chi$ being the spin correlation function and $\Im$ stands for the imaginary part. Owing to the two legs of the ladder, there are 
two channels for spin-spin correlations: even and odd. Spin-spin correlations in the even (odd) channel are calculated with respect to
$\vec{S}_{1} \pm \vec{S}_{2}$. It is straightforward to see that within the single-mode approximation (no continuum contribution) 
only the odd channel is relevant. After a few steps of algebra we obtain
\begin{widetext}
\begin{align}
\label{eq:sx_def}
S^{xx}_{ij} &= \sum^{N}_{m=1} \bigg\lbrace \delta(\omega - \omega_{mx}) 
\big[ - v^{*}_{i,m} u^{*}_{j,m} + v^{*}_{i,m} v_{j,m} %\nonumber \\ 
%&~~~~~~~
+ u_{i,m} u^{*}_{j,m} - u_{i,m} v_{j,m} \big] \nonumber \\
&~~~~~~~~+\delta(\omega - \omega_{mz}) 
\big[ - v^{*}_{i,m+N} u^{*}_{j,m+N} + v^{*}_{i,m+N} v_{j,m+N} %\nonumber \\ 
%&~~~~~~~
+ u_{i,m+N} u^{*}_{j,m+N} - u_{i,m+N} v_{j,m+N} \big]  \bigg\rbrace \,, \\
%%%%%%%%%%
\label{eq:sz_def}
S^{zz}_{ij} &= \sum^{N}_{m=1} \bigg\lbrace \delta(\omega - \omega_{mx}) 
\big[ - v^{*}_{i+N,m} u^{*}_{j+N,m} + v^{*}_{i+N,m} v_{j+N,m} %\nonumber \\ 
%&~~~~~~~
+ u_{i+N,m} u^{*}_{j+N,m} - u_{i+N,m} v_{j+N,m} \big] \nonumber \\
&~~~~~~~~+\delta(\omega - \omega_{mz}) 
\big[ - v^{*}_{i+N,m+N} u^{*}_{j+N,m+N} + v^{*}_{i+N,m+N} v_{j+N,m+N} %\nonumber \\ 
%&~~~~~~~
+ u_{i+N,m+N} u^{*}_{j+N,m+N} - u_{i+N,m+N} v_{j+N,m+N} \big]  \bigg\rbrace  \,.
\end{align}
\end{widetext}
The structure factor $S(\omega)$ is then obtained by summing all the spin contributions (here, $x$ and $z$). 
This is plotted in Fig. (\ref{fig:s_w}). Note that here we have neglected broadening of the lines and so there are only sharp delta peaks. Inside the topological paramagnet, we can see a dominant sharp peak at the end-state energy with smaller peaks (much weaker intensity) from the bulk triplon bands. In the trivial phase there is no such peak at the end-state energy.

%%%%%%%%%%%%%%%%%%%%%%%%%%%%%%%%%%%

%%%%%%%%%%%%%%%%%%%%%%%%%%%%%%%%%%%
\section{Alternative model}

\begin{figure}
\centering
\includegraphics[width=0.4\textwidth]{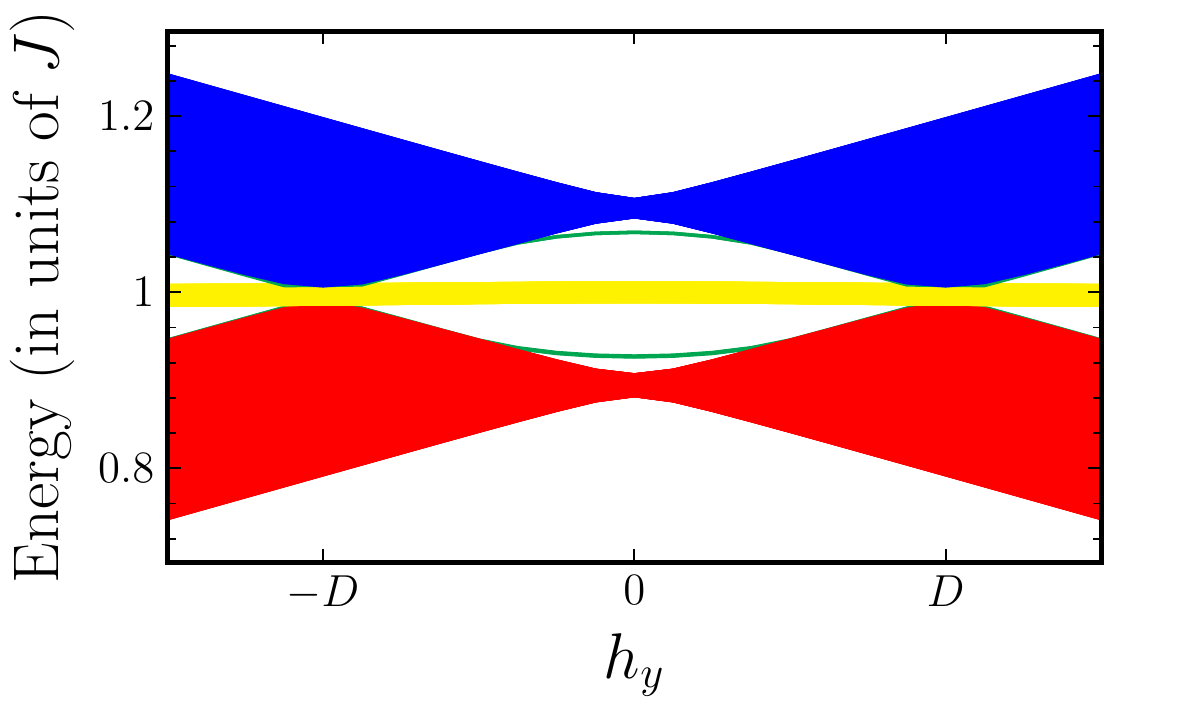}
\caption{Band structure for the alternative model, Eq.~\eqref{eq:ham_har_n}, with open ends. Protected end states  (green line) appear in the topological paramagnetic phase, $|h_y|<D$. In contrast to the model discussed in the main text, the end-state energy varies with $\hy$ and the end states merge with the bulk well before $\hy=|\dm|$. The parameters used here are $\dm/\jp=\gm/\jp=0.1$ and $\jn/\jp=0.01$.
}
\label{fig:es_new}
\end{figure}

%{\clg 
In the main text we have considered a model where the even-parity term $\Gamma$ points in the same direction as the DM interaction.
However, the direction of the $\gm$ term is not fixed by lattice symmetries. It can also point in some other direction,  depending on the arrangement of the atoms in a given compound.  But the topological phase
is expected to be present for any direction of the $\gm$ term. Here we show this, by considering a model where
the $\gm$ term points along the $z$ direction.
The corresponding Hamiltonian is given by %}
\begin{eqnarray}
\mathcal{H} 
&&= 
\jp \sum_{i} \vec{S}_{1i} \cdot \vec{S}_{2i} 
+ \jn \sum_{i} \big[ \vec{S}_{1i} \cdot \vec{S}_{1i+1} + \vec{S}_{2i} \cdot \vec{S}_{2i+1} \big] \nonumber \\
&&+ \dm \sum_{i} \big[ S^{z}_{1i} S^{x}_{1i+1} - S^{x}_{1i} S^{z}_{1i+1} 
+ S^{z}_{2i} S^{x}_{2i+1} - S^{x}_{2i} S^{z}_{2i+1} \big] \nonumber \\
&&+ \gm \sum_{i} \big[ S^{x}_{1i} S^{y}_{1i+1} + S^{y}_{1i} S^{x}_{1i+1} 
+ S^{x}_{2i} S^{y}_{2i+1} + S^{y}_{2i} S^{x}_{2i+1} \big] \nonumber \\
&&+ \hy \sum_{i} \big[ S^{y}_{1i} + S^{y}_{2i} \big] \,.
\label{eq:model_new}
\end{eqnarray}
The main difference between the above model and that considered in the main text [Eq. (1)] is that, here the even-parity spin anisotropic interaction $\gm$ points in different direction compared to that of the odd-parity DM interaction. Using Eq.~\eqref{eq:s_op} and \eqref{eq:s1s2}, we obtain the corresponding triplon Hamiltonian:
\begin{align}
\mathcal{H} &= -\frac{3}{4} \jp N + \jp \sum_{i \alpha} t^{\dagger}_{i\alpha} t_{i\alpha}  \nonumber \\
&+ \frac{\jn}{2} \sum_{i\alpha} \big[ t^{\dagger}_{i\alpha} P_{i} P_{i+1} t_{i+1 \alpha}
- t^{\dagger}_{i\alpha} P_{i} t^{\dagger}_{i+1\alpha} P_{i+1} + H.c. \big] \nonumber \\
&+\frac{\dm}{2} \sum_{i} \big[ t^{\dagger}_{iz} P_{i} P_{i+1} t_{i+1 x} - t^{\dagger}_{iz} P_{i} t^{\dagger}_{i+1x} P_{i+1}  
\nonumber \\
&~~~~~~~~~~- t^{\dagger}_{ix} P_{i} P_{i+1} t_{i+1 z} + t^{\dagger}_{ix} P_{i} t^{\dagger}_{i+1z} P_{i+1} + H.c. \big] \nonumber \\
&+\frac{\gm}{2} \sum_{i} \big[ t^{\dagger}_{ix} P_{i} P_{i+1} t_{i+1 y} - t^{\dagger}_{ix} P_{i} t^{\dagger}_{i+1y} P_{i+1} 
\nonumber \\
&~~~~~~~~~~+ t^{\dagger}_{iy} P_{i} P_{i+1} t_{i+1 x} - t^{\dagger}_{iy} P_{i} t^{\dagger}_{i+1x} P_{i+1} + H.c. \big] \nonumber \\
&+\ii \hy \sum_{i} \big[ t^{\dagger}_{ix} t_{iz} - t^{\dagger}_{iz} t_{ix} \big] \,.
\label{eq:ham_r_n}
\end{align}
%{\clg 
Keeping only the bilinear piece, which amounts to the harmonic approximation, we obtain %}
\begin{align}
\mathcal{H}_{2} &=  \jp \sum_{i \alpha} t^{\dagger}_{i\alpha} t_{i\alpha}  %\nonumber \\
+ \frac{\jn}{2} \sum_{i\alpha} \big[ t^{\dagger}_{i\alpha} t_{i+1 \alpha}
- t^{\dagger}_{i\alpha} t^{\dagger}_{i+1\alpha}  + H.c. \big] \nonumber \\
&+\frac{\dm}{2} \sum_{i} \big[ t^{\dagger}_{iz} t_{i+1 x} - t^{\dagger}_{iz} t^{\dagger}_{i+1x}  
- t^{\dagger}_{ix} t_{i+1 z} + t^{\dagger}_{ix} t^{\dagger}_{i+1z}  + H.c. \big] \nonumber \\
&+\frac{\gm}{2} \sum_{i} \big[ t^{\dagger}_{ix} t_{i+1 y} - t^{\dagger}_{ix} t^{\dagger}_{i+1y}  
+ t^{\dagger}_{iy} t_{i+1 x} - t^{\dagger}_{iy} t^{\dagger}_{i+1x} + H.c. \big] \nonumber \\
&+\ii \hy \sum_{i} \big[ t^{\dagger}_{ix} t_{iz} - t^{\dagger}_{iz} t_{ix} \big] \,.
\label{eq:ham_har_n}
\end{align}
In contrast to the model described in the main text, here all the three triplon modes are coupled even at the harmonic level. Even in this case, the quantum paramagnetic phase is split into a topological and a trivial part. The topological phase transition occurring again at $\hy=|\dm|$. The eigenmodes of the bilinear Hamiltonian [Eq.~\eqref{eq:ham_har_n}] with open boundary condition are plotted in 
Fig.~\ref{fig:es_new}. Apart from the three bulk triplon modes, we also find doubly degenerate end states at each end of the ladder in the topological phase. Interestingly, the end-state energy changes as a function of $\hy$ and the end states merge with the bulk triplon bands even before the bulk bands touch each other at $\hy=|\dm|$.

%%%%%
%%%%%

In this case, we could not find a suitable transformation which will bring the bilinear Hamiltonian [Eq.~\eqref{eq:ham_har_n}], 
written in momentum space, in an off-block diagonal form. So an expression for the winding number can not be derived. However, we can make some progress by considering only the hopping piece. For small $\gm$ and $\dm$ the effect of anomalous terms is negligible, such that the full bilinear Hamiltonian is adiabatically connected to the hopping Hamiltonian. The hopping piece in the momentum space is given by
\begin{subequations} \label{triplon_ham_k_new}
\begin{equation}
\mathcal{H}_{k,hop} = \frac{1}{2} \sum_{k} \Psi^{\dagger}_{k} \mathcal{H}_{1 k} \Psi_{k} ,
\label{eq:ham_k}
\end{equation}
with the spinor
$
\label{eq:psi}
\Psi_{k} = 
(
t_{kx} , t_{ky}, t_{kz} 
)^{T} $
%%%%%%
and the $2 \times 2$ matrix
\begin{equation}
H_{1} (k) = 
(\jp + \jn \cos (k)) \mathbbm{1} + \vec{d'} \cdot \vec{L} \,,
\end{equation}
such that 
\begin{equation}
\vec{L} =  \left\lbrace
\begin{bmatrix}
0 & 1 & 0 \\
1 & 0 & 0 \\
0 & 0 & 0 
\end{bmatrix} \,, 
%%%%
%L_{2} = 
\begin{bmatrix}
0 & 0 & \ii \\
0 & 0 & 0 \\
-\ii & 0 & 0 
\end{bmatrix} \,, 
%%%%
%L_{3} = 
\begin{bmatrix}
0 & 0 & 0 \\
0 & 0 & 1 \\
0 & 1 & 0 
\end{bmatrix} 
\right\rbrace
\end{equation}
\end{subequations}
satisfy the SU(2) algebra: $\big[ L_{i}, L_{j} \big]=\ii \epsilon_{ijk} L_{k}$, and 
$\vec{d'}=\left\lbrace \gm \cos (k), \dm \sin (k) +\hy, 0 \right\rbrace$. The winding number corresponding to 
$\mathcal{H}_{k,hop}$ is easy to evaluate,
\begin{equation}
\mathcal{W}=\frac{1}{2 \pi} \bigg[ \int \frac{1}{|d'|^2} \left( d'_{1} d(d'_{2}) - d'_{2} d(d'_{1}) \right) \bigg] \,.
\end{equation}
It is non-zero in the topological quantum paramagnetic phase, $\hy<|\dm|$, while it is zero in the trivial phase. It is therefore reasonable to assume that even upon including the anomalous terms the topological phase exists. This is also evident from Fig.~\ref{fig:es_new}, where we can clearly see end states even in the presence of the anomalous terms.

%%%%%%%%%%%%%%%%%%%%%%%%%%%%%%%%%%%

 %\bibliographystyle{apsrev}
%\bibliography{bib_topo_para_supp_v1}

%%%%%%%%%%%%%%%%%%%%%%%%%%%%%%%%%%%%%%%%%%%%%%%%%%%%%%%%%%%%%%%%%%%%%%%%%%%%%%%
%

%%%%%%%%%%%%%%%%%%%%%%%%%%%%%%%%%%%%%%%%%%%%%%%%%%%
\end{document}